\documentclass{pramana}

\usepackage{graphicx,bm}
\usepackage{multirow}%
\usepackage{amsmath,amssymb,amsfonts}%
\usepackage{mathrsfs}%
\usepackage[title]{appendix}%
\usepackage{xcolor}%
\usepackage{textcomp}%
\usepackage{manyfoot}%
\usepackage{booktabs}%
\usepackage{algorithm}%
\usepackage{algorithmicx}%
\usepackage{algpseudocode}%
\usepackage{listings}%
\usepackage{caption}%
\usepackage{hyperref}
\usepackage[labelformat=simple]{subcaption}

\begin{document}

%%paper title
%%For line breaks \\ can be used within title 
\title{Pattern Formation in Excitable Neuronal Maps}

%%author names are separated by comma (,) 
%%use \and before the last author name 
%%\textsuperscript{number} is used for affiliation
%%use a * along with the number separated by comma
%% for the  author for correspondence

\author{Divya D. Joshi, Trupti R. Sharma\and Prashant M. Gade\textsuperscript{*}}
\affilOne{Department of Physics, Rashtrasant Tukadoji 
 Maharaj Nagpur University, Amravati Road, Nagpur, India- 440033}
%%escape two column mode for title, affiliation and abstract
%%by giving \twocolumn command as shown

\twocolumn[{

\maketitle

%%include \corres to print the corresponding author Email id
\corres{prashant.m.gade@gmail.com}

%%include \msinfo for
%%manuscript information such as
%%received, revised and accepted dates
%%
%\msinfo{1 January 2015}{1 January 2015}{1 January 2015}

%%abstract
\begin{abstract} 
Coupled excitable systems can generate a variety of patterns. In this work, we investigate coupled Chialvo maps in two dimensions under two types of nearest-neighbor couplings. One coupling produces ring-like patterns, while the other produces spirals. The rings expand with increasing coupling, whereas spirals evolve into turbulence and dissipate at stronger coupling. To quantify these patterns, we introduce an analogue of the discriminant of the velocity gradient tensor and examine the persistence of its sign. For ring-type patterns, the persistence decays more slowly than exponentially, often following a power law or stretched exponential. When spiral structures remain intact, persistence saturates asymptotically and can exhibit superposed periodic oscillations, suggesting complex exponents at early times. These behaviors highlight deep connections with the underlying dynamics.
\end{abstract}

%%insert keywords separated by comma using \keywords{words}
\keywords{Coupled maps, Pattern formation, Chialvo map, spiral, ring}

%%include \pacs{number} to print the PACS number
%\pacs{12.60.Jv; 12.10.Dm; 98.80.Cq; 11.30.Hv}

}]
%%close the twocolumn escape here

%%include \doinum{number}for the DOI number in the header
%%include \volnum{number} for the volume number in the header
%%include \year{yyyy} for  year of publication in the header
%%include \pgrange{num--num} page range of article in the header
%%include \artcitid{num} for the article citation id
%%include \lp to print last page of the article
%%include \setcounter{page}{pagenum} for the exact starting page of the article

%\doinum{12.3456/s78910-011-012-3}
%\artcitid{\#\#\#\#}
%\volnum{123}
%\year{2016}
%\pgrange{23--25}
%\setcounter{page}{23}
%\lp{25}

\section{Introduction}\label{sec1}
Excitable media provide another pathway to pattern formation through the propagation of self-generated waves. This behaviour is characterized by the existence of a threshold of activation.  Small fluctuations in the resting state of the system rapidly decay without having an effect on the system. Only strong stimulus, {\it{i.e.}}, one which exceeds the threshold, is able to make the system switch to another state distinct from the rest state, termed the \textquoteleft activated' or \textquoteleft excited' state. Once an element in this system reaches its excited state, it stays there only for a finite duration before slowly returning to its equilibrium resting state within a time interval known as the \textquoteleft refractory period'. Wave patterns formed by excitable media play a very important role in physical, chemical, or biological processes. In excitable media, the wave pattern formed by self-organization represents a broad and intensively developed field of study in nonlinear dynamical systems and has important potential applications, {\it{e.g.}} in cardiology \cite{app1,app2,app3,app4}.

Specifically, the phenomenon of patterns of excitation is a characteristic feature of spatially extended excitable media. Spatial patterns in an excitable system arise from a distinct property of mutual annihilation on collision of interacting excitation waves. These spatial patterns are referred to variously as \textquoteleft reentrant excitations'(1-D), \textquoteleft vortices or spiral waves'(2-D), and \textquoteleft scroll waves'(3-D) \cite{sitabhra}. In particular, the dynamics of spiral waves can be described accurately by closely related excitable models. Under certain circumstances, the singularity defining the vortex of the spiral wave can itself drift in space over time. If at any time this meandering motion brings it close to another region of the spiral, the two wavefronts collide and mutually annihilate. This results in multiple free ends of the wavefronts, each of which eventually curls around, forming several coexisting spiral waves. Thus, instabilities in the spiral waves can lead to a state characterized by many coexisting small spirals, which is a manifestation of spatiotemporal chaos often termed turbulence \cite{guo}. Such a state has not only been observed in real systems but has also been associated with clinical conditions, such as arrhythmias, which are pathological deviations from the normal rhythmic activity of the heart. A certain type of arrhythmia, tachycardia, during which heart tissue is activated at an abnormally rapid rate, is related to the genesis of spiral waves. It was realized that spiral breakup and subsequent irregular activity correspond to cardiac fibrillation, during which the heart effectively stops pumping blood and can be fatal unless controlled within minutes \cite{sinha}. Thus, the studies in formation and control of these patterns have well-defined applications.

Rigidly rotating spirals, critical fingers, and wave segments are 
commonly moving patterns in excitable 2-D media. But the most investigated 
are spiral waves, used as a paradigmatic example of  
self-organization \cite{app3}. Wave segments are unstable, but can be stabilized by applying appropriate feedback to excitable media \cite{waveseg}. They define a separatix between spiral wave behaviour and contracting wave segments \cite{separ}.
Unbounded wave segments(critical fingers) lie on the asymptote 
of the separatix, defining the boundary between excitable and 
subexcitable media \cite{karma,zykov}.

Excitable media show self-sustained wave propagation of various geometries. The excitation pulse may be circulated along a closed one-dimensional ring. Most theoretical interest has focused on the study of spiral waves and scroll waves in excitable media. However, there are fewer observations of the formation of expanding rings in a 2D medium.  If a stable pulsating  spot or a stationary
spot appears in 1D, we can observe an expanding or stationary ring
in two dimensions\cite{vasiev2004classification}. Such patterns have been observed experimentally\cite{budrene1995dynamics,poptani2022biological}. This is considered an extension of travelling waves in 1D systems\cite{morozov2009excitable}. Expanding rings are susceptible to noise if the velocity is small\cite{vasiev2004classification} and transform into labyrinthine patterns.  We shall not consider the impact of noise in this work. It has been argued that scroll rings can be conveniently created from expanding circular sings\cite{mikhailov2006control}. Such expanding rings in an excitable medium are also studied in \cite{fraser1986ring} using a simple automaton model for dynamics. Different automaton models of expanding rings are reviewed and compared in \cite{gravner1996cellular}.
%but there is not yet a good mathematical theory for the relatively simple problem of pulse circulation on a ring. This problem has concrete applications to experimental models of reentrant electrical activity in cardiac muscle \cite{cm}. In 1914, Mines considered the circulation of an electrical pulse (or action potential) around a ring-shaped piece of atrial muscle dissected from a tortoise heart. He proposed this preparation as a model for abnormal reentrant activity \cite{mines}. An experiment by Frame and Simson \cite{FS} on ring-shaped myocardial preparations dissected around the tricuspid valve orifice of dog hearts has provided detailed new results on the dynamics of pulse circulation in rings of living cardiac tissue. They found that, in certain preparations, the magnitude of these quantities oscillated when measured once each rotation at a fixed point along the ring of tissue and that such oscillations often accompanied the termination of reentrant propagation around the ring. Hence, an understanding of the nature of these oscillations could be helpful clinically in controlling the stability of anatomical reentry circuits. These are excitable systems, and a model of one-dimensional local dynamics is not useful for describing them. We study a system of coupled two-dimensional maps to model and analyze the properties of these systems.

Pattern formation in one-dimensional coupled map lattices has often been quantified using quantifiers such as
synchronization error, flip rate, or persistence \cite{gade2006,pakhare2020A,gade2007}. The underlying map is also one-dimensional, and the reference is the fixed point. However, for two-dimensional systems, a single point cannot divide the phase space into two parts, and a new quantifier is needed. Taking inspiration from a quantifier 
introduced in the context of turbulence \cite{mitra}, we introduce a discretized version of the Okubo-Weiss order parameter. 

\section{Model}\label{sec2}
Along with continuous-time models, for the study of neural
activity, discrete time models have been recently used. Among discrete models, the system suggested by Chialvo \cite{chi} is well known. The system proposed by Chialvo is,

\begin{eqnarray}
 x_{n+1}&=&f(x_n,y_n)=x^2_{n}\exp({y_n-x_n})+k,\nonumber\\
 y_{n+1}&=&g(x_n,y_n)=ay_n-bx_n+c.
 \label{eq1}
\end{eqnarray}
where, $x_{n}$ acts as activation variable and $y_{n}$ as 
a recovery-like variable. Subscript $n$ represents iteration steps. The model includes four parameters. The parameter $k$ acts as a constant bias or as a time-dependent additive perturbation. For $x=0$ the fixed point of the recovery variable is determined by three parameters: $a$, the time constant of recovery $(a<1)$; $b$, the activation-dependence of the recovery process $(b<1)$, and offset $c$. The map is rich in dynamics, from oscillatory to chaotic. We study coupled map lattice where dynamics on each site are defined by an excitable map introduced by Chialvo. The coupled map lattice  with two different types of coupling is defined as: \\
{\bf{Nonlinear coupling:}}
\begin{eqnarray}
 x_{t+1}^{i,j}&=&(1-\epsilon) f(x_{t}^{i,j},y_{t}^{i,j})\nonumber \\
 &&+{\frac{\epsilon}{4}} ((f(x_{t}^{i-1,j},y_{t}^{i-1,j})+f(x_{t}^{i,j-1},y_{t}^{i,j-1})\nonumber\\
 &&+f(x_{t}^{i+1,j},y_{t}^{i+1,j})+f(x_{t}^{i,j+1},y_{t}^{i,j+1})),\nonumber\\
 y_{t+1}^{i,j}&=&g(x_{t}^{i,j},y_{t}^{i,j}).
 \label{eq2}
\end{eqnarray}
We also study strongly nonlinear coupling given below.\\
{\bf{Nonlinear quadratic coupling:}}
\begin{eqnarray}
 x_{t+1}^{i,j}&=&(1-\epsilon) f(x_{t}^{i,j})
 %+\epsilon/4\ (f(x_{t}^{i-1,j})+f(x_{t}^{i,j-1})+f(x_{t}^{i+1,j})+f(x_{t}^{i,j+1}))^{2},\nonumber\\
 % y_{t+1}^{i,j}&=&g(x_{t}^{i,j}).
 \nonumber \\
 &&+{\frac{\epsilon}{4}} ((f(x_{t}^{i-1,j},y_{t}^{i-1,j})+f(x_{t}^{i,j-1},y_{t}^{i,j-1})\nonumber\\
 &&+f(x_{t}^{i+1,j},y_{t}^{i+1,j})+f(x_{t}^{i,j+1},y_{t}^{i,j+1}))^2,\nonumber\\
 y_{t+1}^{i,j}&=&g(x_{t}^{i,j},y_{t}^{i,j}).
 \label{eq3}
\end{eqnarray}
In the above definitions,  $i$ and $j$ represent the index of each neuron in a square lattice, and $t$ is the time-step.
We define a few more quantities as;
\begin{eqnarray}
 d_i(i,j)\vert_t&=&g(x_t^{i+1,j},y_t^{i+1,j})-g(x_t^{i,j},y_t^{i,j})\nonumber\\
 d_j(i,j)\vert_t&=&g(x_t^{i,j+1},y_t^{i,j+1})-g(x_t^{i,j},y_t^{i,j})\nonumber\\
 d_{ii}(i,j)\vert_t&=&(d_i(i+1,j)\vert_t-d_i(i-1,j))\vert_t \nonumber\\
 d_{ji}(i,j)\vert_t&=&(d_i(i,j+1)\vert_t-d_i(i,j-1))\vert_t\nonumber\\
 d_{ij}(i,j)\vert_t&=&(d_j(i+1,j)\vert_t-d_j(i-1,j))\vert_t\nonumber\\
 d_{jj}(i,j)\vert_t&=&(d_j(i,j+1)\vert_t-d_j(i,j-1))\vert_t\nonumber\\
 \label{eq5n}
\end{eqnarray}
where $d$ is the analogue of velocity gradient tensor \cite{mitra}. We define 
discriminant of $d$ as an analog of the Okubo-Weiss parameter:
\begin{eqnarray}
\Gamma_t^{i,j}=((d_{ii}(i,j)+d_{jj}(i,j))^2 -4(d_{ii}(i,j) d_{jj}(i,j)\nonumber \\
-d_{ji}(i,j) d_{ij}(i,j))\vert_t.
 \label{eq5}
\end{eqnarray}

We investigate  plots of the field of $x(i,j)$ for various values of 
$\epsilon$. We studied its persistence for different couplings for $\Gamma_(i,j)$. If
$\Gamma_0(i,j)>0$ and $\Gamma_t(i,j)>0$ for all $t\le T$, it is
persistent till time $T$. Similarly
If $\Gamma_0(i,j)<0$ and $\Gamma_t(i,j)<0$ for all $t\le T$, it is
persistent till time $T$. In other words, the fraction
of sites $(i,j)$ that have not changed the sign of 
$\Gamma^{i,j}$ even once till time $T$ is persistence at time $T$.
Obviously, this is a monotonically decreasing function.
For the Chialvo map, in the case of nonlinear coupling, the variation of $\Gamma_t(i,j)$ shows a ring pattern and persistence decay as a power law for an intermediate range of couplings. For nonlinear quadratic coupling, pattern is spiral and persistence decays as stretched exponentially with asymptotic saturation and 
could decay as a stretched exponential for larger coupling.
\begin{figure*}[]
 \centering
 \begin{tabular}{c}
  \includegraphics[width=0.35\textwidth]{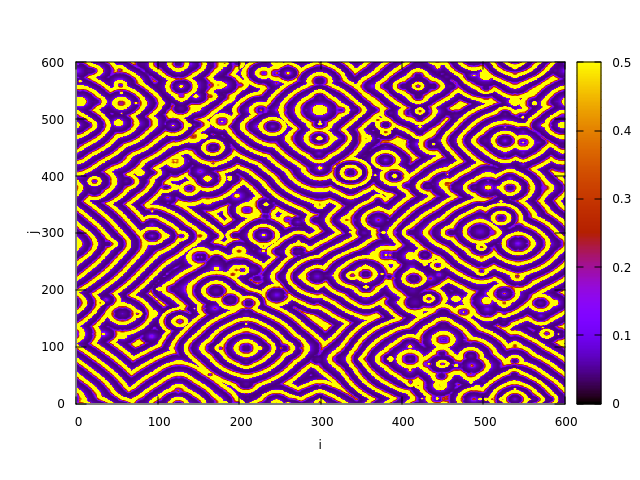}
  \small(a)
 \end{tabular}
 \begin{tabular}{c}
  \includegraphics[width=0.35\textwidth]{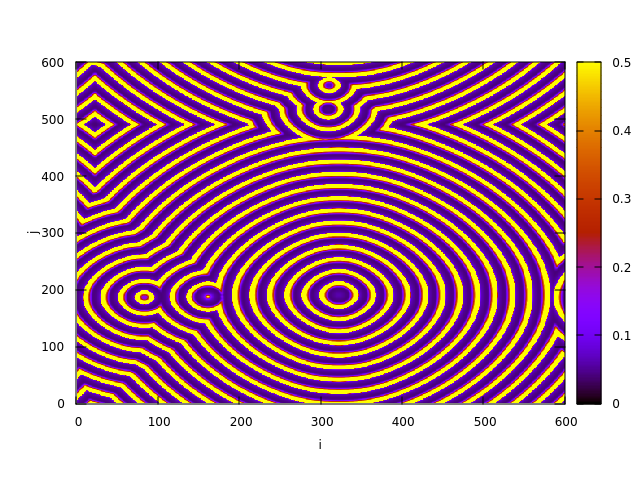}
  \small(b)
 \end{tabular}
 \begin{tabular}{c}
  \includegraphics[width=0.35\textwidth]{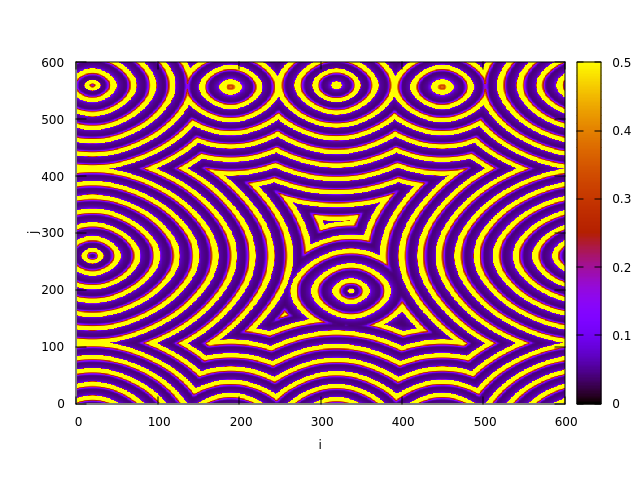}
  \small(c)
 \end{tabular}
 \begin{tabular}{c}
  \includegraphics[width=0.35\textwidth]{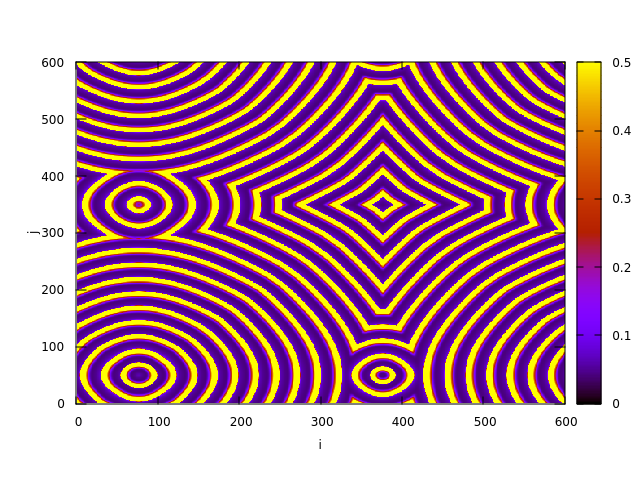}
  \small(d)
 \end{tabular}
 \begin{tabular}{c}
  \includegraphics[width=0.35\textwidth]{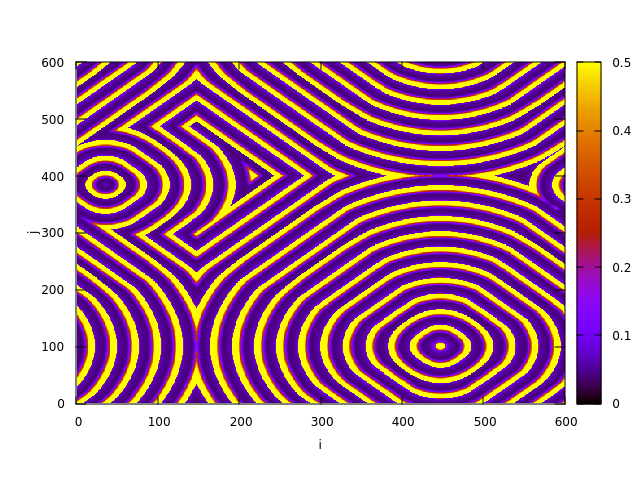}
  \small(e)
 \end{tabular}
 \begin{tabular}{c}
  \includegraphics[width=0.35\textwidth]{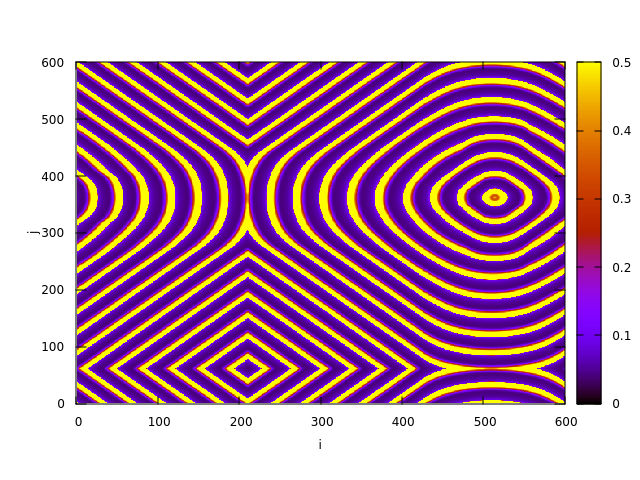}
  \small(f)
 \end{tabular}
 \begin{tabular}{c}
  \includegraphics[width=0.35\textwidth]{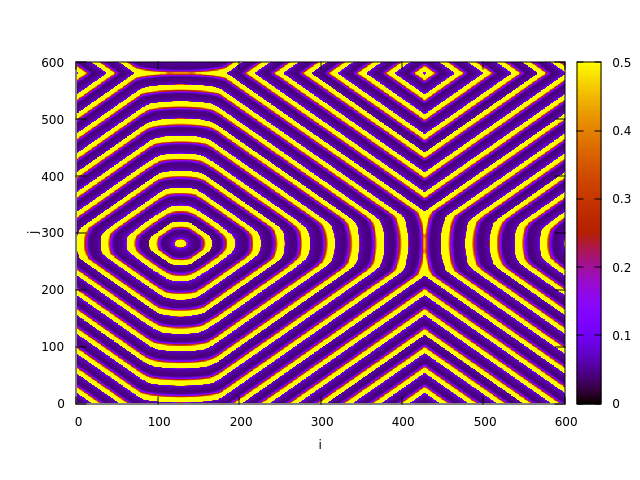}
  \small(g)
 \end{tabular}
 \caption
 {Field of $x(i,j)$ at $t=8\times10^5$ for $\epsilon=0.2,0.3 \dots0.8$ for nonlinear coupling.}
 \label{fig1}%
\end{figure*}
\section{Simulation and Results}\label{sec3}
We simulate the Chialvo map for parameter values $k=0.03, a=0.89, b=0.18$ and $c=0.28$ for two types of coupling $i.e.,$ nonlinear coupling and nonlinear quadratic coupling. We plot the field of $\Gamma$ for various values of $\epsilon$ at $t=10^4$ to observe different patterns and study its persistence for two different couplings.
\subsection{Ring patterns}
\begin{figure*}[]
 \centering
 \begin{tabular}{c}
  \includegraphics[scale=0.2]{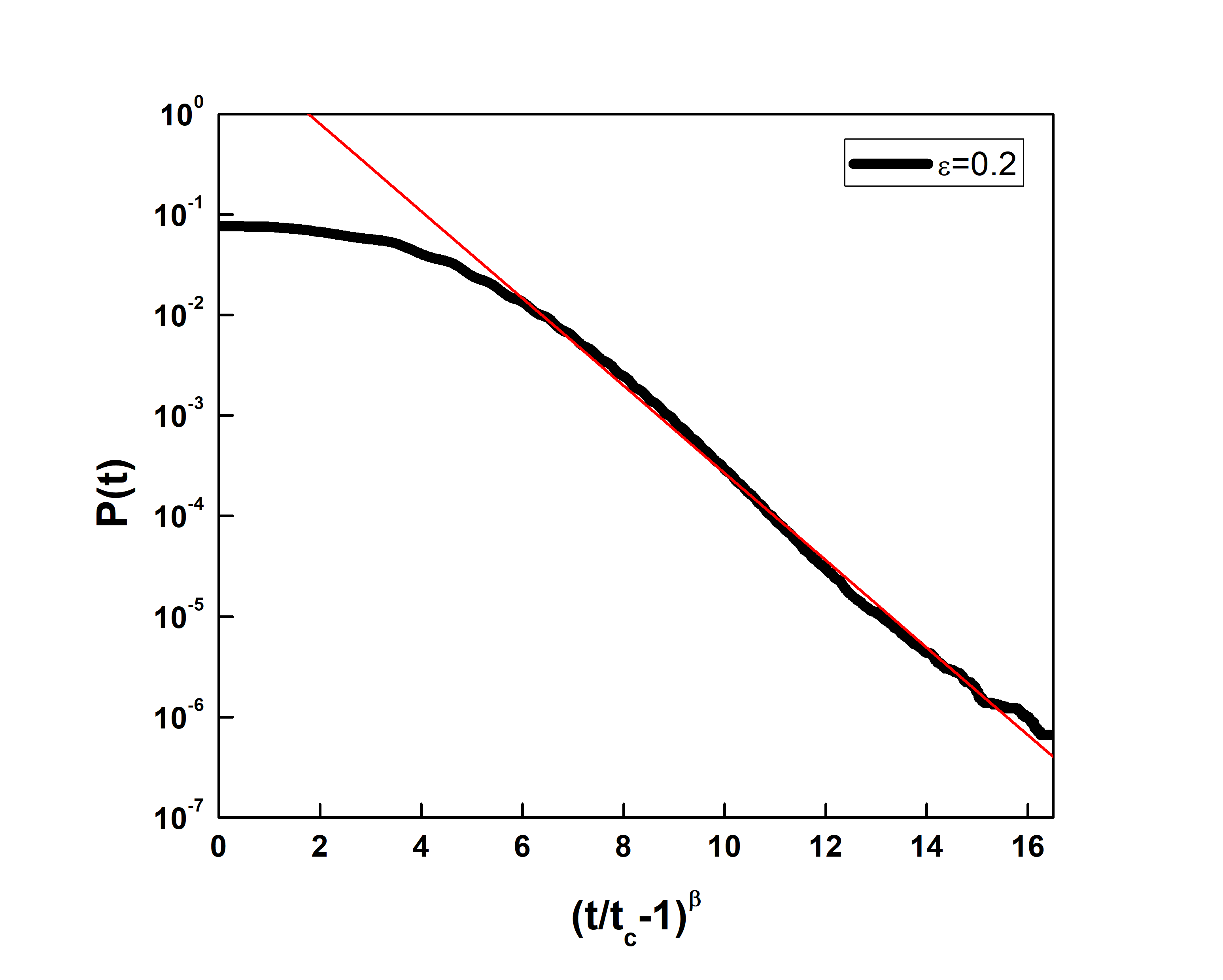}
  \small(a)
 \end{tabular}
 \begin{tabular}{c}
  \includegraphics[scale=0.2]{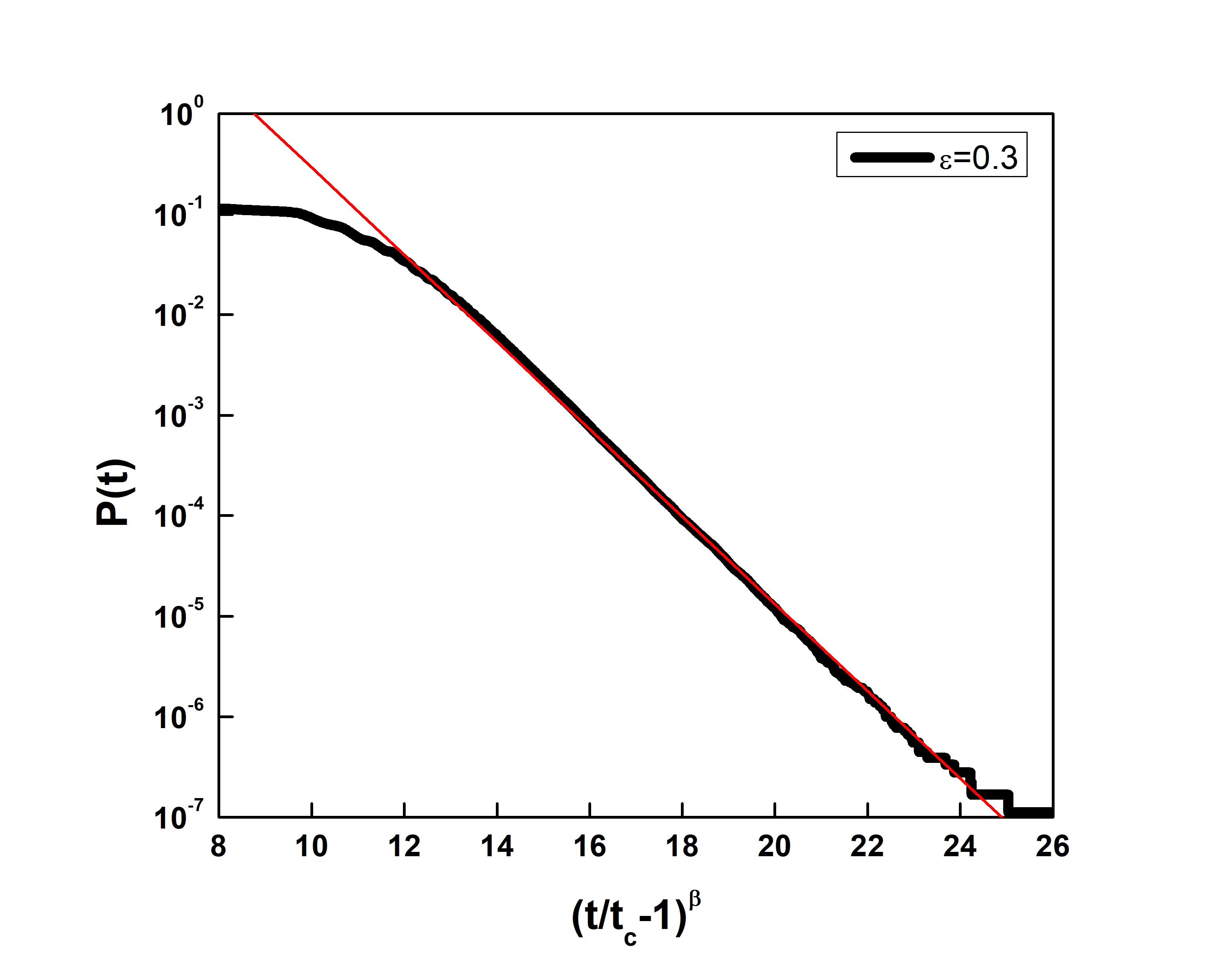}
  \small(b)
 \end{tabular}\\
 \begin{tabular}{c}
  \includegraphics[scale=0.2]{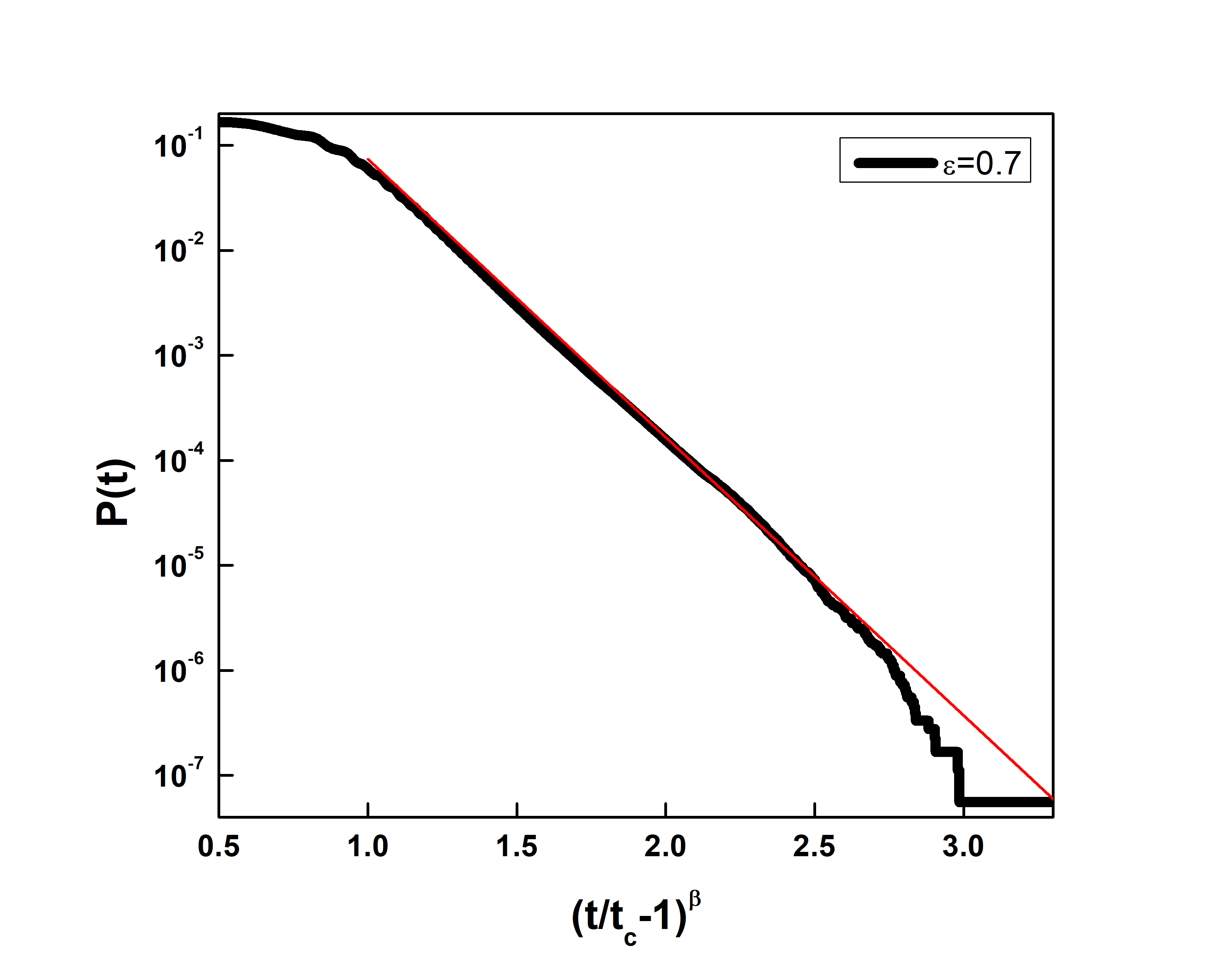}
  \small(c)
 \end{tabular}
 \begin{tabular}{c}
  \includegraphics[scale=0.2]{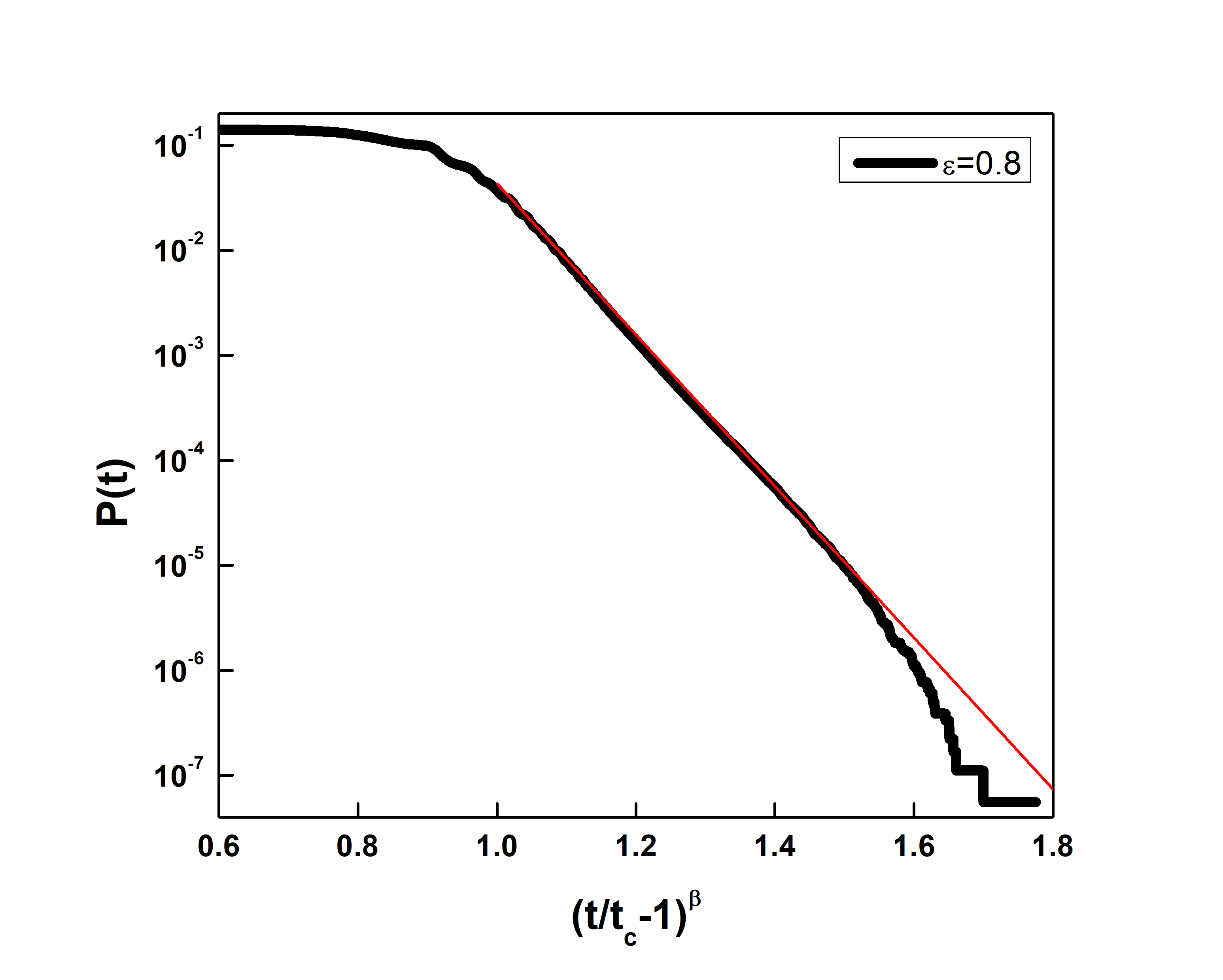}
  \small(d)
 \end{tabular}
 \caption
 {Persistence $P(t)$ as a function of $(t/t_c-1)^\beta$ on a log-normal scale for  $\epsilon=0.2, 0.3, 0.7, 0.8$ with 
  $t_c=125$. It is stretched exponential $P(t)=B\exp(A(t/t_c-1)^\beta)$. An appropriate fit is shown by the red line.  a) For $\epsilon=0.2$, $\beta=0.36$, $A=5.7$ and $B=5.9$. b) For $\epsilon=0.3$, $\beta=0.12$, $A=11.7$ and $B=6500$ c) For  $\epsilon$= 0.7, $\beta=0.2$, $A=6.1$ and $B=33$ d)For $\epsilon$= 0.8, $\beta=0.1$, $A=16.6$ and $B=700000$.}
 \label{fig2}
\end{figure*}

For nonlinear coupling, the field of $x(i,j)$ for various values of $\epsilon$ at $t=10^4$ shows a ring pattern (Fig. \ref{fig1}). As we increase $\epsilon$ from $0.2$ to $0.8$, the rings observed at the same time are bigger in size. No pattern is observed for lower values $\epsilon=0.05$ and $0.1$. For $0.2 \le \epsilon \le 0.8$, rings become fewer and bigger in size.  In each case, we studied the variation of persistence $P(t)$ as a function of time $t$. For a range of values of $\epsilon$, persistence shows power-law behaviour (Fig. \ref{fig2}).  Let $P_+(t)$ be fraction of persistent  sites such that $\Gamma_0(i,j) >0$ and $P_-(t)$ be fraction of persistent  sites such that $\Gamma_0(i,j) <0$. Clearly,
$P(t)=P_+(t)+P_-(t)$. We observe that $P_-(t)$ drops to zero very rapidly, and the behavior of $P(t)$ is dominated by $P_+(t)$.

The decay of persistence is slower than exponential in all cases. It is stretched exponential for smaller and larger values of $\epsilon$. For an intermediate range of
$\epsilon$, we observe a power-law decay. We have demonstrated  stretched exponential decay
by plotting $(t/t_c-1)^{\beta}$ versus $P(t)$ on semilogarithmic scale in Fig.\ref{fig2}
for $\epsilon=0.2, 0.3, 0.7$ and $0.8$. For $\epsilon=0.4, 0.5$ and $0.6$, we have 
plotted $P(t)$ as a function of $t$ on a logarithmic scale. After a certain initial transient,
a clear power-law is observed over 4 decades. In these cases $P(t)\propto t^{-\gamma}$ and 
$\gamma=1.7$  for $\epsilon=0.4$ and $0.5$ and $1.8$ for $\epsilon=0.6$.  The only distinction between the patterns in Fig. \ref{fig1}c), d), and e)  corresponding to the power-law decay is the following. In this case, in some places, the patterns  can be concave in the gap between two ring patterns. On the other hand, the patterns are always convex, and the rings from different centres smoothly overlap in other cases.
%%%FIGURE%%%
\begin{figure*}[]
 \centering
 \scalebox{0.33}{
  \includegraphics{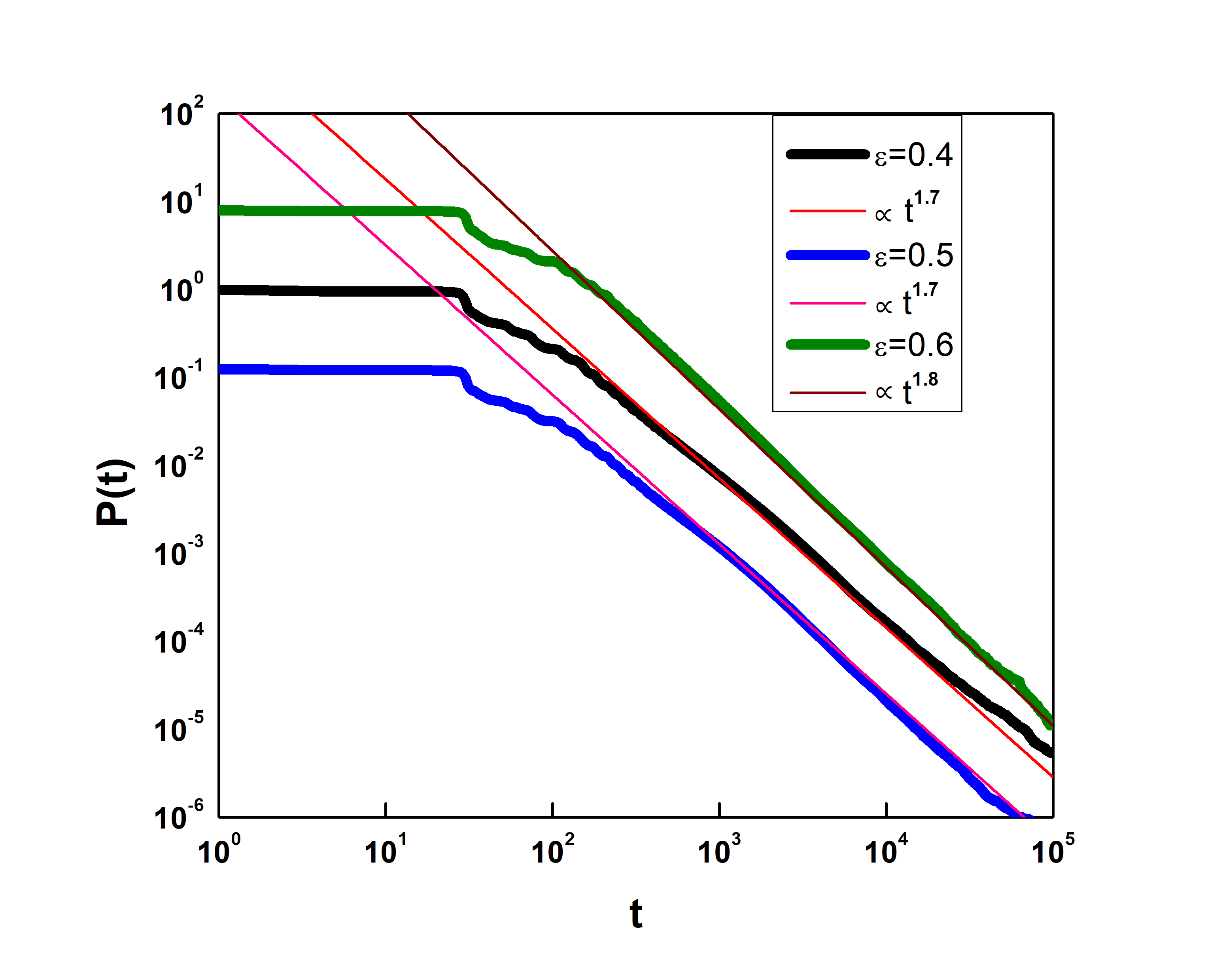}
 }
 \caption{Persistence $P(t)$ as a function of $t$ for $\epsilon=0.4,0.5$ and 0.6 on a log-log scale. The Y-axis is multiplied by arbitrary constants for better visibility. }
 \label{fig2a}
\end{figure*}

\subsection{Spiral patterns}
\begin{figure*}[]
 \centering
 \begin{tabular}{c}
  \includegraphics[width=0.35\textwidth]{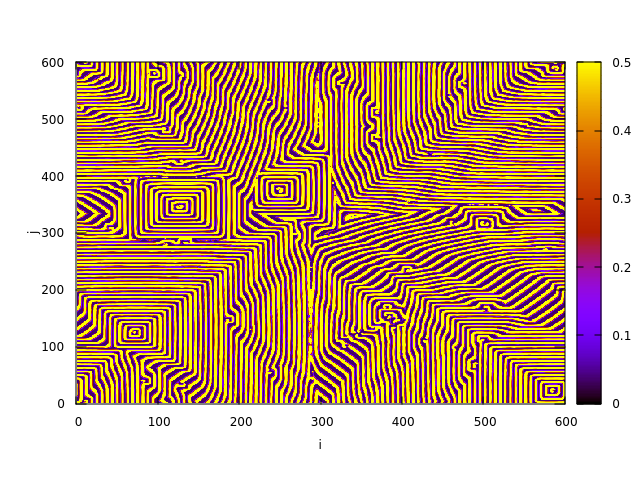}
  \small(a)
 \end{tabular}
 \begin{tabular}{c}
  \includegraphics[width=0.35\textwidth]{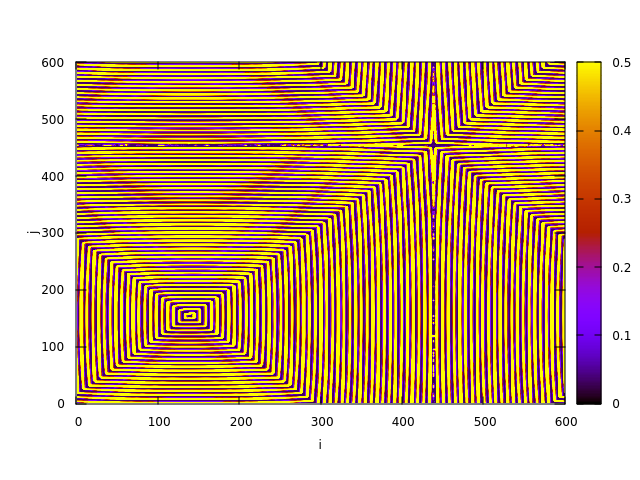}
  \small(b)
 \end{tabular}
 \begin{tabular}{c}
  \includegraphics[width=0.35\textwidth]{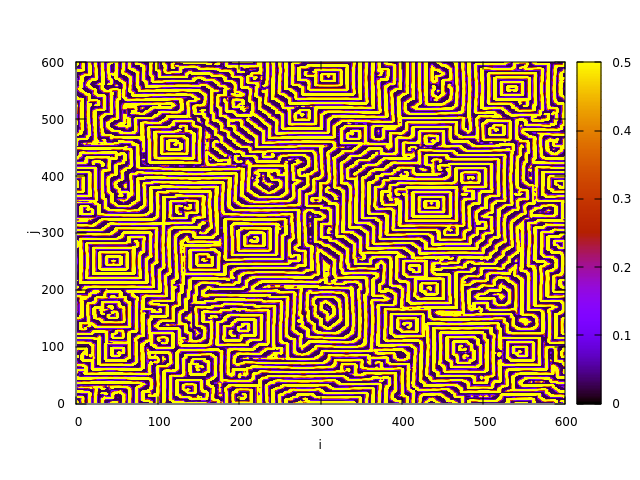}
  \small(c)
 \end{tabular}
 \begin{tabular}{c}
  \includegraphics[width=0.35\textwidth]{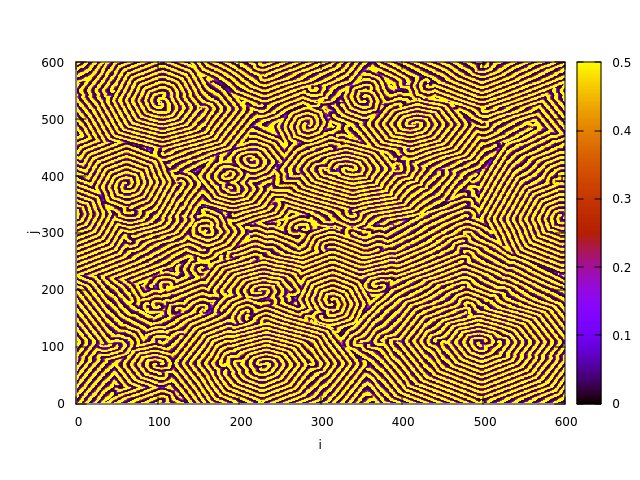}
  \small(d)
 \end{tabular}
 \begin{tabular}{c}
  \includegraphics[width=0.35\textwidth]{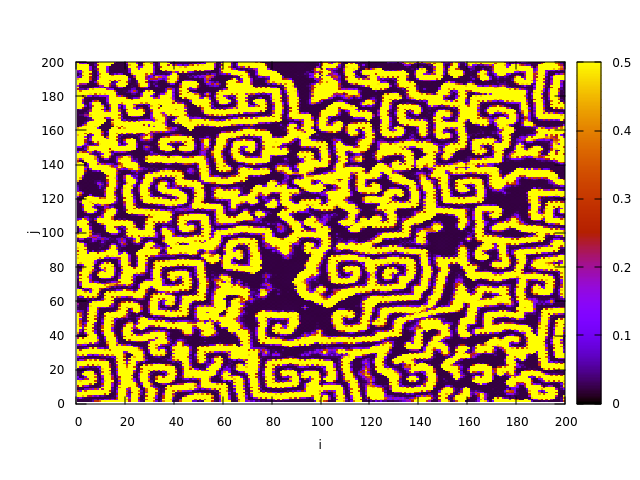}
  \small(e)
 \end{tabular}
 \begin{tabular}{c}
  \includegraphics[width=0.35\textwidth]{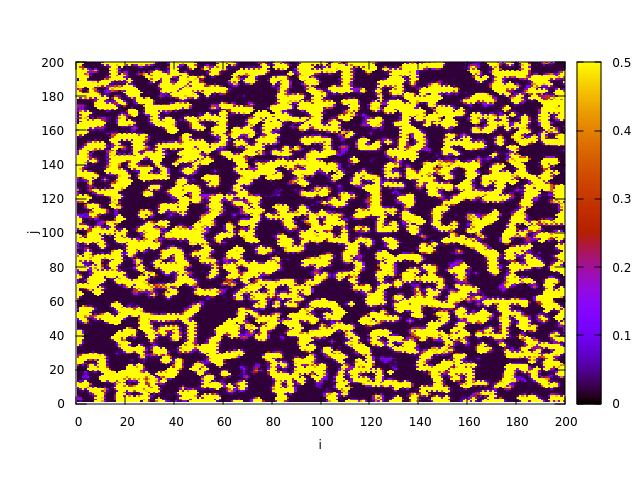}
  \small(f)
 \end{tabular}
 \begin{tabular}{c}
  \includegraphics[width=0.35\textwidth]{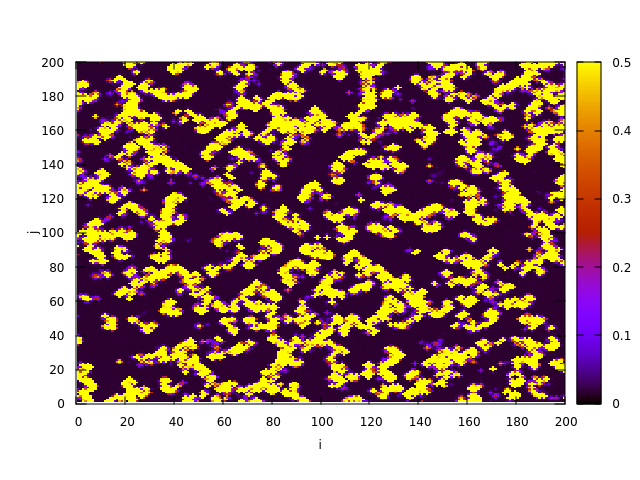}
  \small(g)
 \end{tabular}
 \begin{tabular}{c}
  \includegraphics[width=0.35\textwidth]{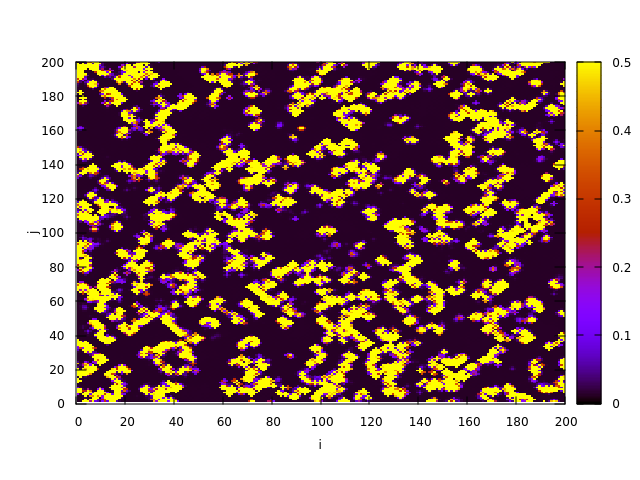}
  \small(h)
 \end{tabular}
 \begin{tabular}{c}
  \includegraphics[width=0.35\textwidth]{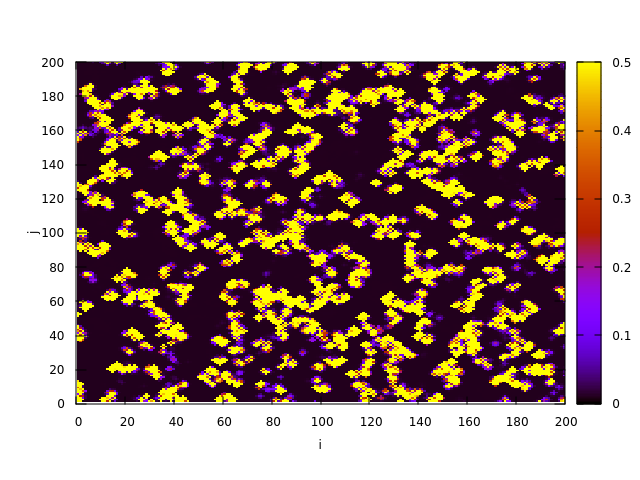}
  \small(i)
 \end{tabular}
 \caption
 {Field of $x(i,t)$ for a) $\epsilon=0.05$, b)$\epsilon=0.1$, c)$\epsilon=0.2$, d)$\epsilon=0.3$, e)$\epsilon=0.4$, e) $\epsilon=0.5$, f) $\epsilon=0.6$ g) $\epsilon=0.7$ and h) $\epsilon=0.8$ for nonlinear quadratic coupling at $t=10^5$. Spirals start breaking for $\epsilon>0.4$}
 \label{fig3}%
\end{figure*}

For nonlinear quadratic coupling, the field of $x(i,j)$ for various values of $\epsilon$ at $t=10^4$, shows a spiral pattern (Fig. \ref{fig3}). As we increase $\epsilon$ from $0.05$ to $0.8$ there is a transition from a stable spiral to the state of turbulence.

\begin{figure*}[]
 \centering
 \begin{tabular}{c}
  \includegraphics[scale=0.24]{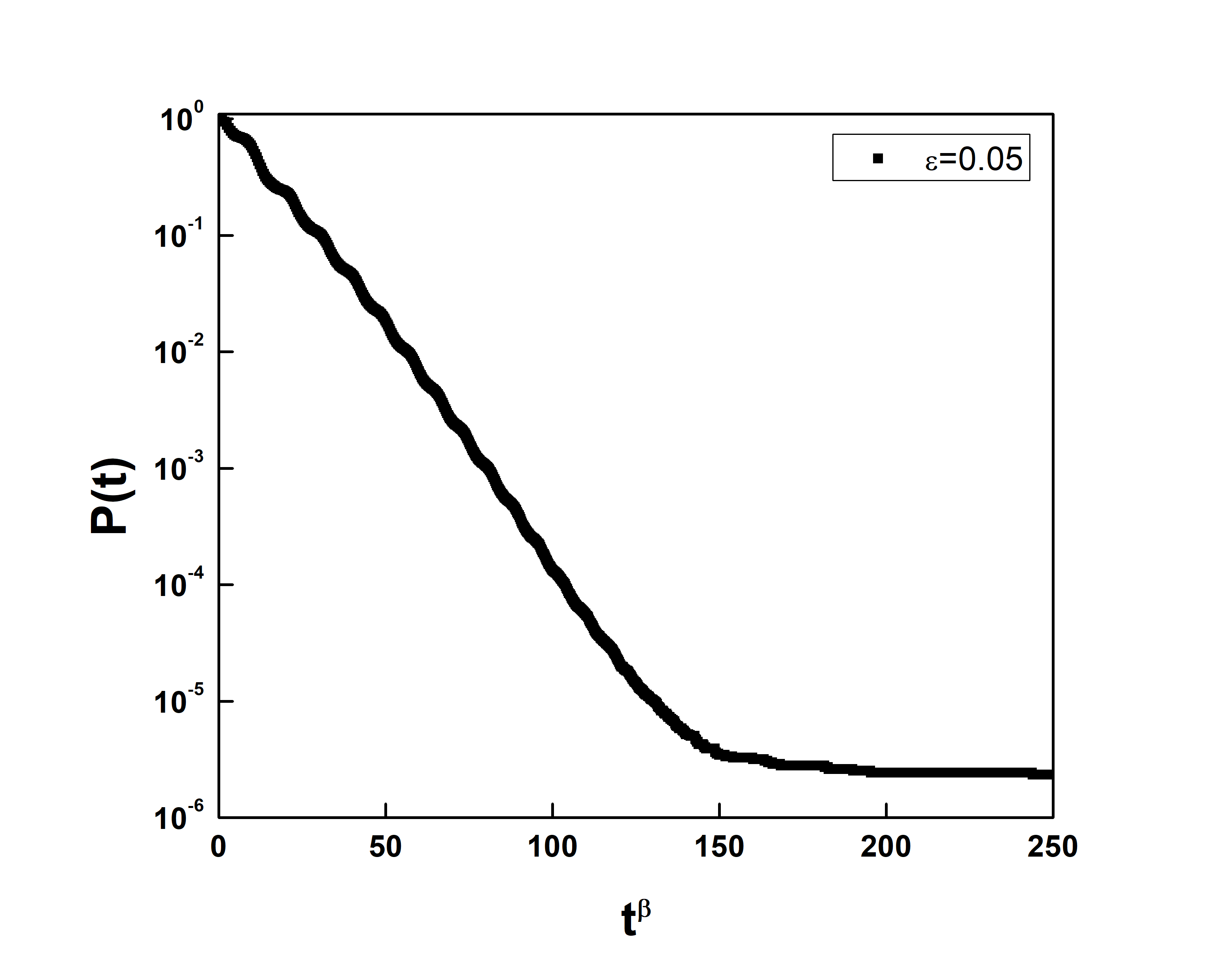}
  \small(a)
 \end{tabular}\\
 \begin{tabular}{c}
  \includegraphics[scale=0.2]{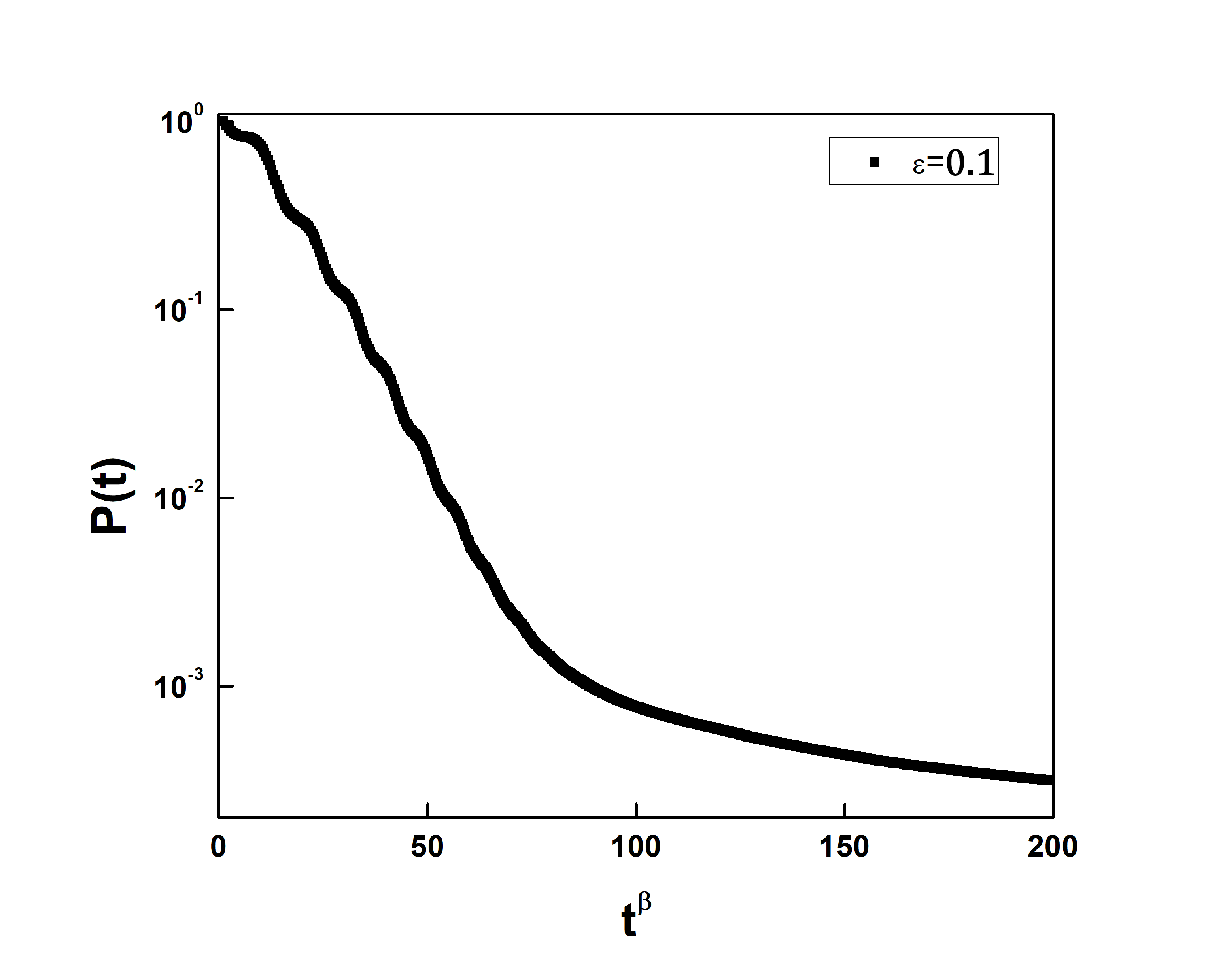}
  \small(b)
 \end{tabular}
 \begin{tabular}{c}
  \includegraphics[scale=0.2]{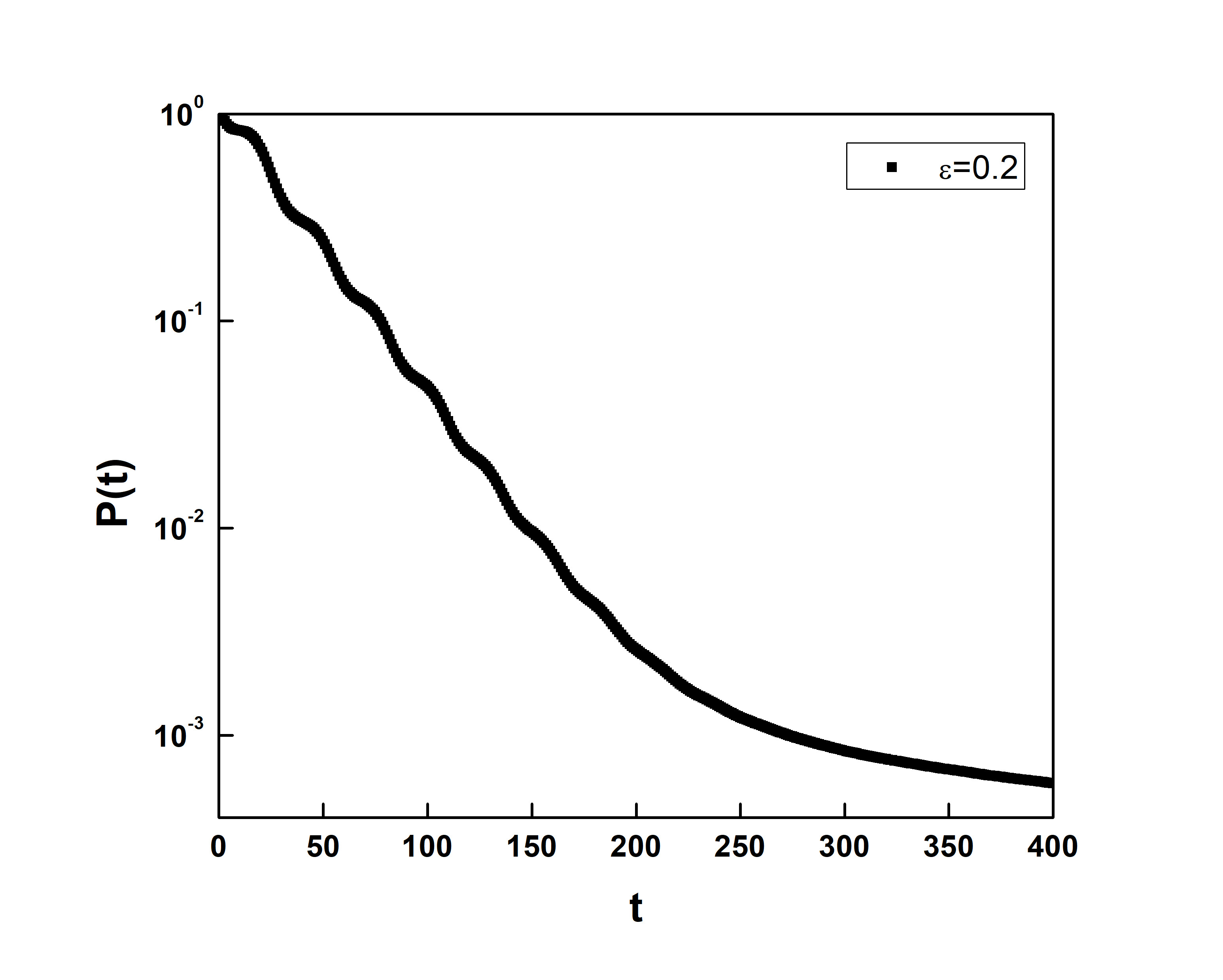}
  \small(c)\\
 \end{tabular}
 \begin{tabular}{c}
  \includegraphics[scale=0.2]{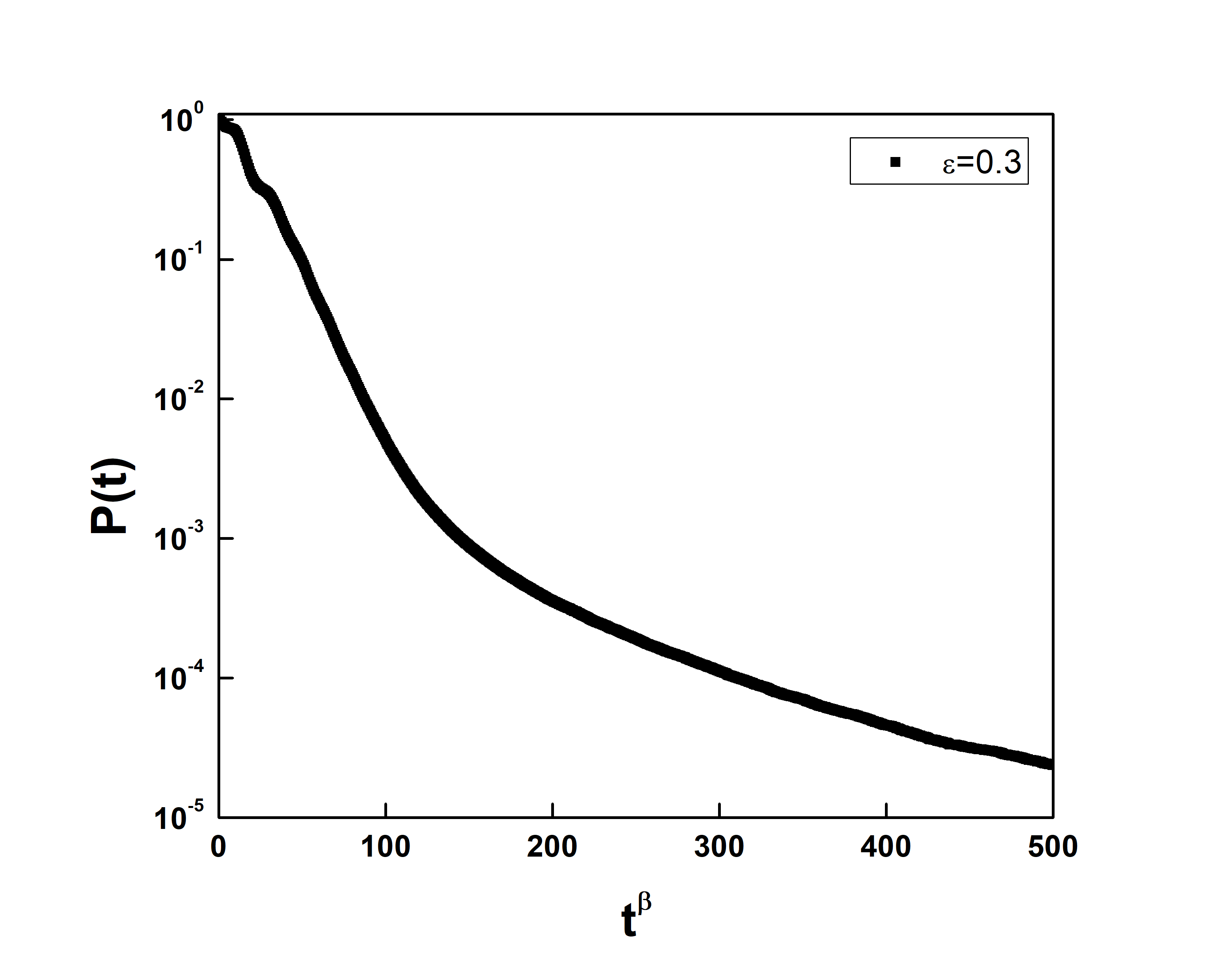}
  \small(d)
 \end{tabular}
 \begin{tabular}{c}
  \includegraphics[scale=0.2]{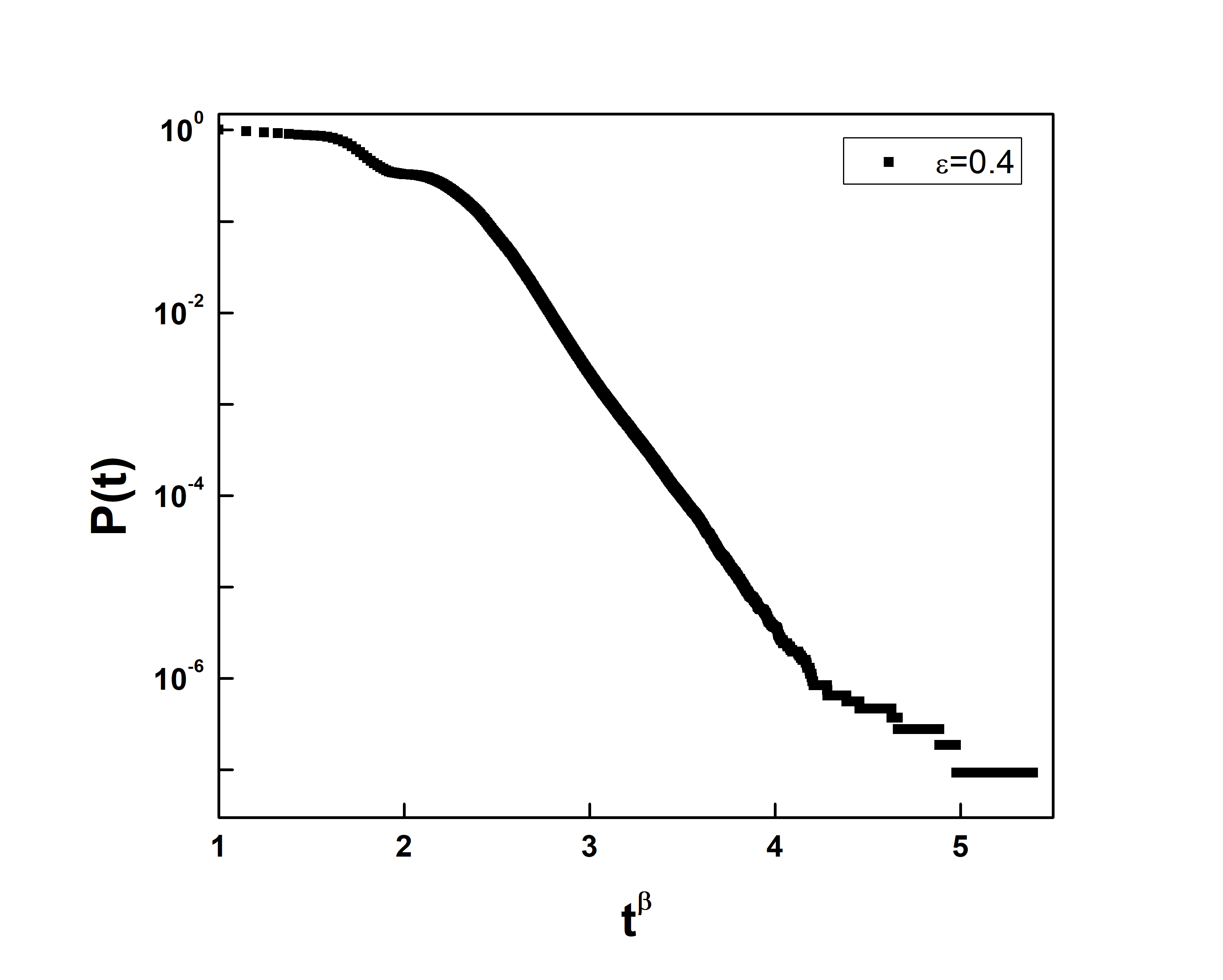}
  \small(e)
 \end{tabular}
 \caption{ Persistence $P(t)$ as a function of $t$ for a) $\epsilon=0.05$ with $\beta=0.8$, b)$\epsilon=0.1$ with $\beta=1$, c)$\epsilon=0.2$ with $\beta=0.8$ d)$\epsilon=0.3$ with $\beta=0.9$, e)$\epsilon=0.4$ with $\beta=0.2$ on semilogarithmic scale. }
 \label{fig4}
\end{figure*}

We plot the persistence, $P(t)$ as a function of time $t$ (Fig. \ref{fig4}). Persistence does not decay as a power law in any case for spiral patterns. Persistence decays as a stretched exponential in all cases {\it{i.e.}} as $\exp(-At^\beta)$. $A$ has dimension of $1/t^\beta$. The values of $\beta$ are 0.8, 1, 0.8, 0.9, 0.2, 0.5, 0.7, and 0.9 for $\epsilon=0.05, 0.1, 0.2, 0.3,0.4,0.5,$ $0.6,0.7,0.8$. For  $\epsilon=0.05,0.1$ and $0.2$ initial decay is superposed by periodic oscillations at early times. It saturates asymptotically for $\epsilon$ ranging from 0.05 to 0.4.  This indicates that certain patterns are frozen and do not move over the time period of observation.  We observe oscillations over and above an exponential for small values of $\epsilon$ only. For larger values of $\epsilon$, it does not saturate. This correlates well with the breakdown of spiral structures for larger values of $\epsilon$. For $\epsilon>0.4$, we observe large laminar regions that allow small spiral structures to move, leading to persistence decay.
\begin{figure*}[]
 \centering
 \begin{tabular}{c}
  \includegraphics[scale=0.2]{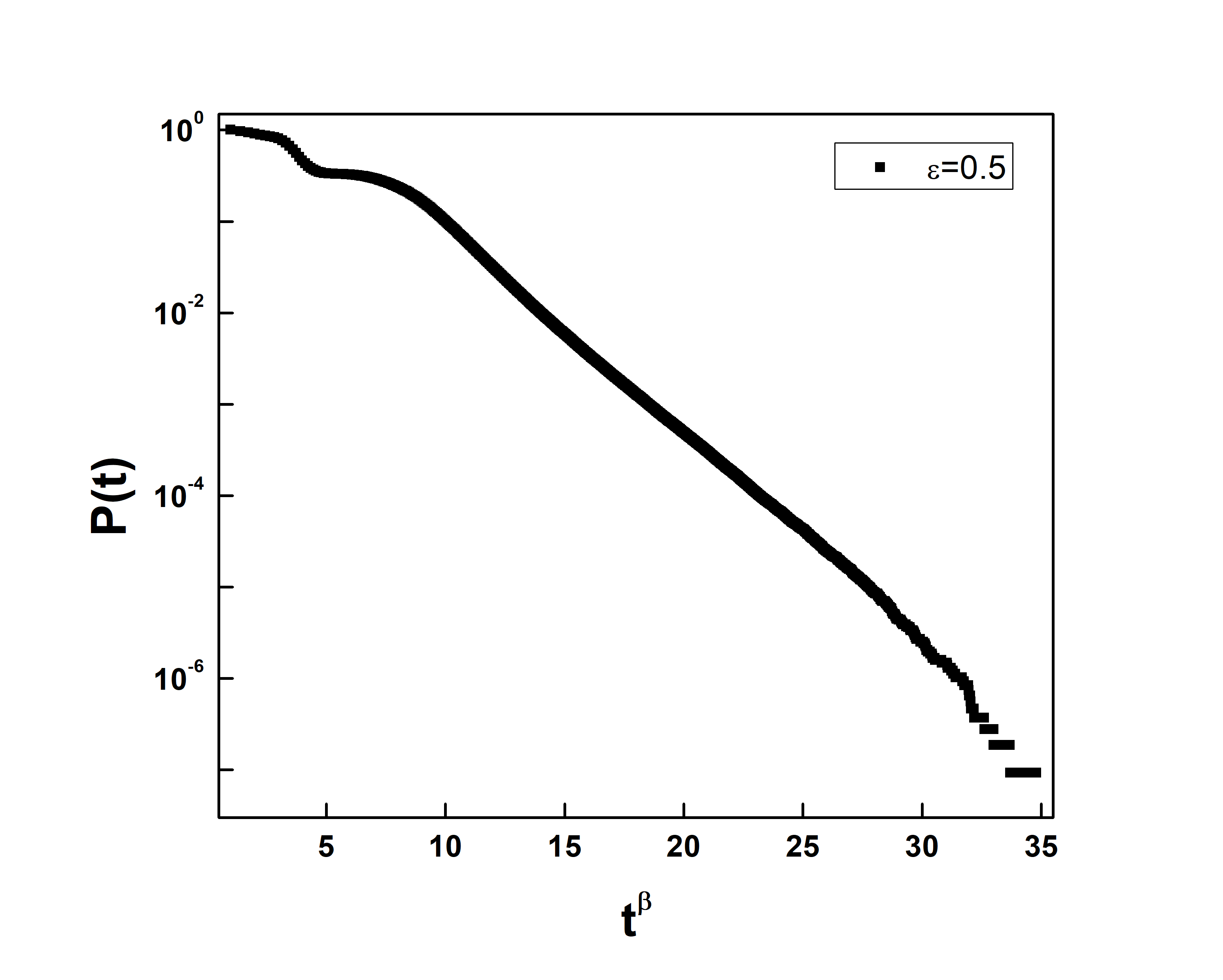}
  \small(a)
 \end{tabular}
 \begin{tabular}{c}
  \includegraphics[scale=0.2]{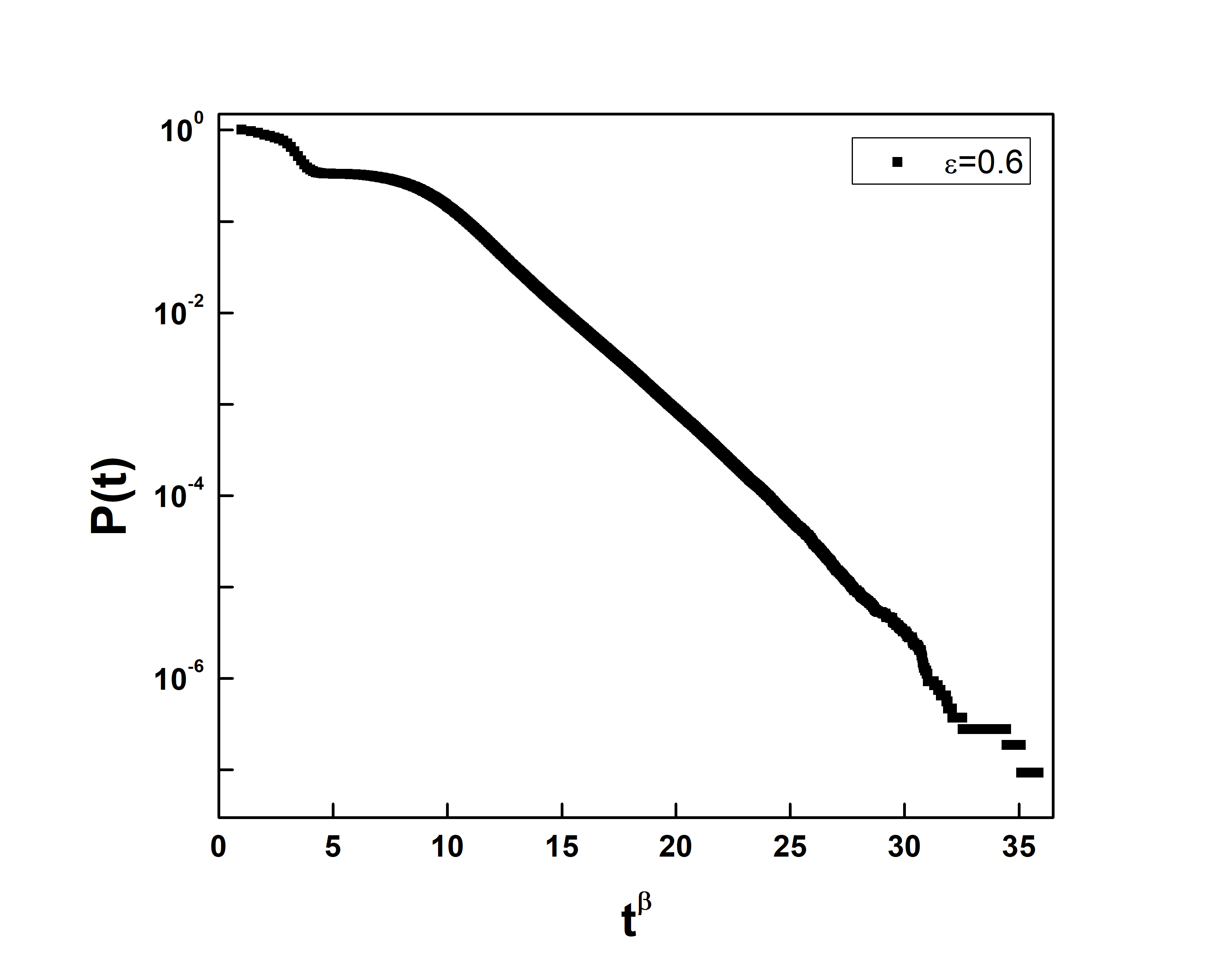}
  \small(b)
 \end{tabular}\\
 \begin{tabular}{c}
  \includegraphics[scale=0.2]{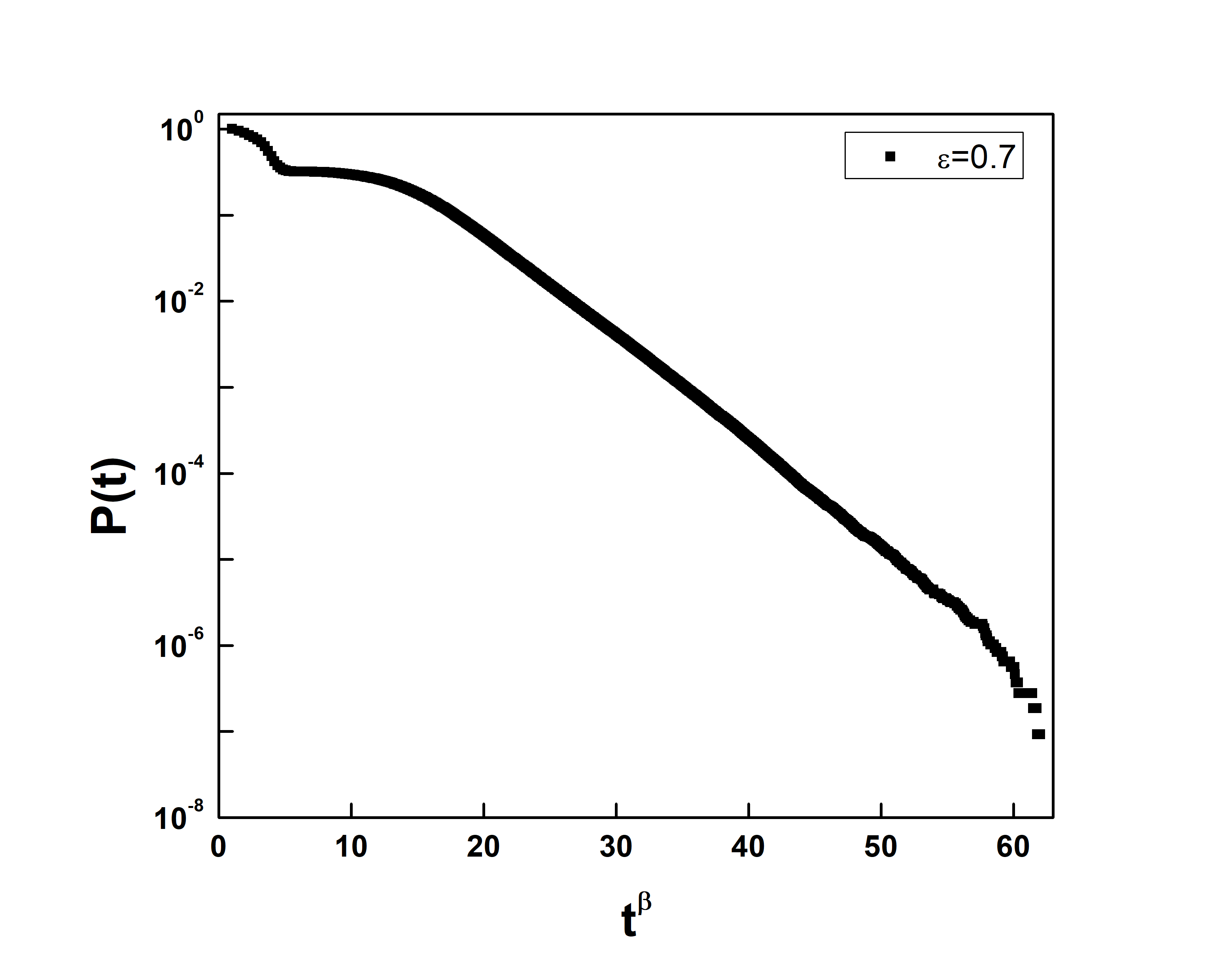}
  \small(c)
 \end{tabular}
 \begin{tabular}{c}
  \includegraphics[scale=0.2]{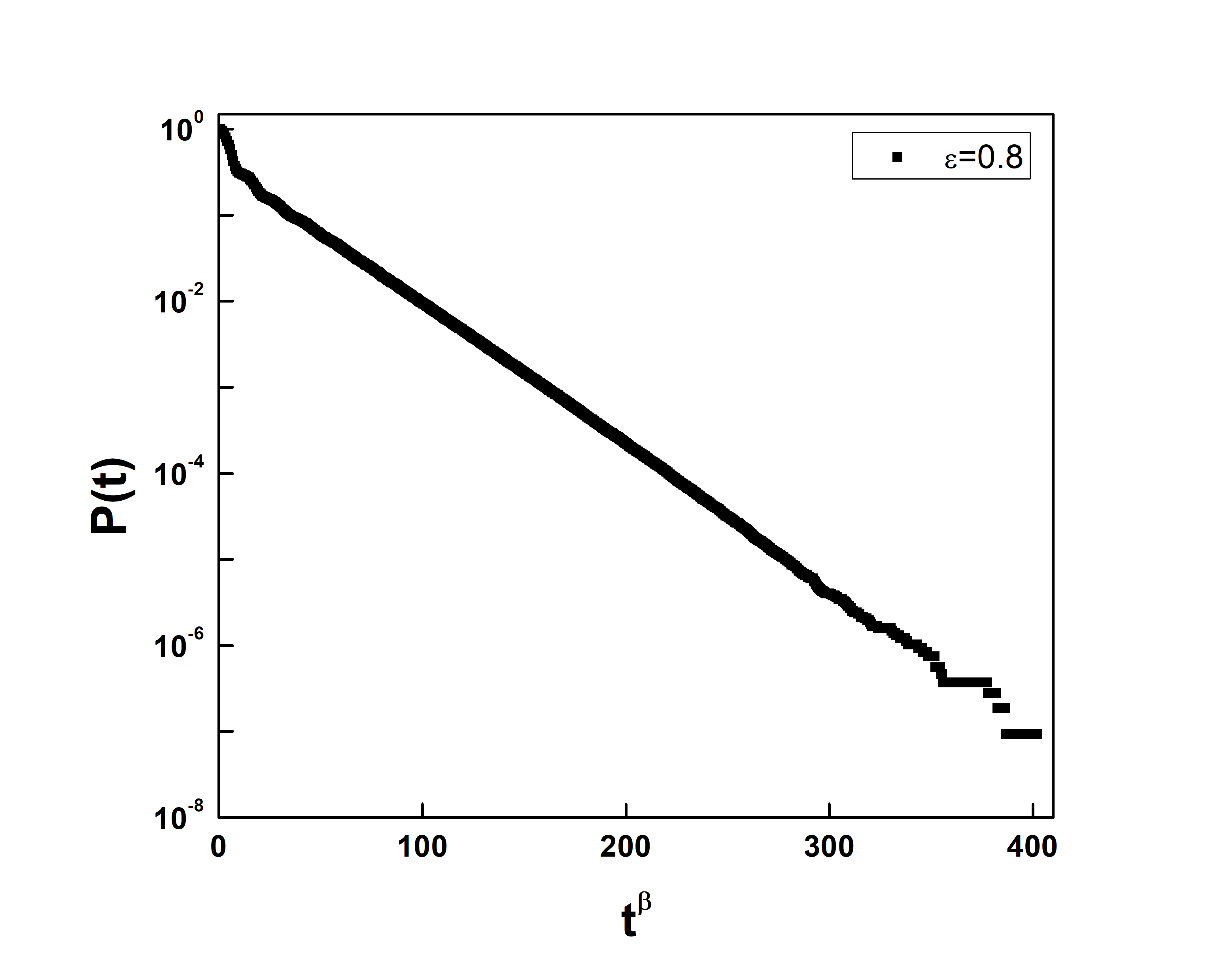}
  \small(d)
 \end{tabular}
 \caption{ Persistence $P(t)$ as a function of $t$ for  a)$\epsilon=0.5$ with $\beta=0.5$, b)$\epsilon=0.6$ with $\beta=0.5$, c)$\epsilon=0.7$ with $\beta=0.7$, and d)$\epsilon=0.8$ with $\beta=0.9$ on semilogarithmic scale. }
 \label{fig5}
\end{figure*}

\section{Discussion}\label{sec12}
Persistence has been defined and studied in one-dimensional coupled map lattices where each map is one-dimensional \cite{menon2003,gade2013}.  Several such studies have been carried out in the context of transition to synchronized or absorbing state, synchronized periodic state, coarse-grained synchronization, as well as
the transition to zigzag pattern \cite{pakhare2020A,gaiki2024,sabe2024synchronization,warambhe,deshmukh2021}. Often, we observe a power-law in persistence at the critical point, and 
the exponents match directed percolation, Ising, or directed Ising class. A new universality class is also observed in
this context \cite{gaiki2024} and is modeled using the cellular automata model \cite{joshi2024cellular}.

Since a single point cannot split the phase space into two in higher dimensions, unlike in a one-dimensional situation, this definition is inapplicable to two-dimensional maps. Therefore, a fresh definition is needed.  There are singularities in the \textquotedblleft flow \textquotedblright, and the singularity may be surrounded by ring or spiral patterns. In the context of turbulence, these singularities have been measured using the Okubo-Weiss parameter inspired by the work in \cite{mitra}. To quantify them, we use the same parameter. In particular, we study sign persistence
of this quantity.

In this study, we examine two different kinds of linked neural maps. Two asymptotic pattern types—ring and spiral—are discernible. We define the discretized version of the Okubo-Weiss parameter and study persistence 
in the sign of this quantity.  For spiral patterns and small coupling, we observe oscillations over exponential decay, indicating a complex exponent. In this case, the pattern is frozen over the observed time period, and the persistence saturates. For higher coupling, the persistence does not saturate and decays as a stretched exponential. When the spiral patterns break, and the laminar region percolates through space, there is space for spiral defects to move, leading to decay of persistence eventually.  For ring patterns, persistence decays slower than exponential for all values of coupling. For an intermediate range of couplings,  a power-law decay of persistence is observed. In this case, the rings originating from different centers do not overlap smoothly at the boundaries. We do not observe any nonzero asymptotic value of persistence for ring patterns. All of the above studies shed light on an intriguing connection with the dynamics of the patterns, how they spread, and how they connect with the persistence of the sign of the discriminant.  It indicates that the time evolution of this parameter can be useful in identifying and 
distinguishing between different patterns and their evolution. 

PMG and DDJ thank SERB grant (CRG/2020/003993) for financial assistance.

\bibliography{sn-bibliography.bib}% common bib file

\end{document}